# A promising approach for the real-time quantification of cytosolic protein-protein interactions in living cells

Ilaria Incaviglia[1], Andreas Frutiger[1], Yves Blickenstorfer[1], Fridolin Treindl[1], Giulia Ammirati[1], Ines Lüchtefeld[1], Birgit Dreier[2], Andreas Plückthun[2], Janos Vörös[1], Andreas M Reichmuth[1]*

1 Laboratory of Biosensors and Bioelectronics, Institute for Biomedical Engineering, ETH Zurich, 8092 Zurich, Switzerland
2 Department of Biochemistry, University of Zurich, 8057 Zurich, Switzerland

*corresponding author: reichmuth@biomed.ee.ethz.ch

## Abstract

In recent years, cell-based assays have been frequently used in molecular interaction analysis. Cell-based assays complement traditional biochemical and biophysical methods, as they allow for molecular interaction analysis, mode of action studies and even drug screening processes to be performed under physiologically relevant conditions. In most cellular assays, biomolecules are usually labeled to achieve specificity. In order to overcome some of the drawbacks associated with label-based assays, we have recently introduced 'cell-based molography' as a biosensor for the analysis of specific molecular interactions involving native membrane receptors in living cells. Here, we expand this assay to cytosolic protein-protein interactions. First, we created a biomimetic membrane receptor by tethering one cytosolic interaction partner to the plasma membrane. The artificial construct is then coherently arranged into a two-dimensional pattern within the cytosol of living cells. Thanks to the molographic sensor, the specific interactions between the coherently arranged protein and its endogenous interaction partners become visible in real-time without the use of a fluorescent label. This method turns out to be an important extension of cell-based molography because it expands the range of interactions that can be analyzed by molography to those in the cytosol of living cells.

# Introduction

Molecular interactions between cellular proteins are commonly referred to as protein-protein interactions (PPIs) and are of central importance to every biological system[1–3]. PPIs play an active role in virtually all cellular functions including proliferation, apoptosis, cell metabolism, angiogenesis, metastasis and immune destruction. Even though the significance of PPIs is widely recognized, high-affinity drugs that selectively modulate or disrupt the interaction between cytosolic proteins are still relatively rare compared to other drug classes[4]. This can be attributed to two main reasons. Firstly, the contact surfaces involved in intracellular PPIs are typically devoid of deep grooves, clefts or pockets that act as binding sites for natural ligands and could be easily exploited for the development of high affinity small molecule drugs[5]. This makes the discovery of novel compounds inherently challenging. Secondly, there is a lack of suitable technologies for PPI analysis that work inside living cells. This is due to the fact that many cytosolic PPIs lack enzymatic functions related to their interaction, which is prohibitive for the development of broadly applicable assays[6]. Consequently, this latter point is worth expanding on.

Since the past years, biophysical and biochemical techniques including fluorescence-based methods, NMR spectroscopy, X-ray crystallography and surface plasmon resonance have been successfully used for the identification and analysis of fundamental interactions as well as for drug screening[7–11]. These techniques are essential to establish specificity for the protein of interest; however, most of them fail at providing a physiologically relevant context, which is crucial for the understanding of cellular processes and mode of action of drugs. In this respect, live cell assays provide information needed to elucidate the effects of drug candidates in a physiological environment. However, they introduce the problem that the observed effect can no longer be exclusively attributed to the protein of interest.

To minimize this problem, several live cell assays rely on genetically encoded fluorescent labels to track the movement of single proteins, enable kinetic analysis of PPIs and for high-throughput screening of modulating compounds[12–14]. Nonetheless, label-based cellular assays have some drawbacks. For instance, resonant energy transfer (RET) biosensors necessitate for the fluorescent reporters to be in close proximity (<10 nm apart), which may require biomolecular optimization and makes it difficult to quantify PPIs[15,16]. Furthermore, they require that one or even both of the reaction partners are modified and recombinantly expressed, which could significantly perturb their function and interaction dynamics.

Recently, label-free live cell biosensors have emerged as an attractive alternative to overcome some of these drawbacks. These sensors measure the bulk refractive index or impedance changes caused by adherent cells on a sensor surface. Any redistribution of cellular content within the first few hundred nanometers above the sensor surface results in an overall change of the respective bulk property. This is then used to infer morphological changes and phenotypic cellular responses. Though incredibly powerful, label-free live cell methods fail to discern specific molecular interactions and are inherently difficult to interpret, as they can only provide information about cytosolic mass, solute concentration and volume changes[17–20].

Here we introduce a generalizable, quantitative live cell assay for the analysis of cytosolic PPIs, which is based on the working principle of focal molography[21–23]. Focal molography is an analytical method for molecular interaction analysis that is insensitive to non-specific molecular interactions. This unique property is achieved by the arrangement of binding sites into a regular (*i.e.*, coherent) pattern, termed 'mologram' (Fig. 1a). The molographic pattern alternates ridges, which are usually made to contain high-affinity probes, and non-functional grooves, which act as an intrinsic reference. Binding of target analytes to the probes on the ridges results in the physical manifestation of a diffraction grating in the shape of the mologram. The mologram is designed such that the light of a guided laser beam is scattered at specifically bound molecules and constructively interferes in a diffraction-limited focal spot[21,23]. The light intensity in the focal spot is measured to quantify molecular binding in real-time. Meanwhile, the light scattered by randomly bound molecules does not contribute to this effect.

Based on this principle, we previously introduced cell-based molography, where specific molecular interactions between membrane proteins and cytosolic proteins are measured in living cells[24]. In cell-based molography, the molographic pattern is established within the membrane of living cells (Fig. 1b). This is achieved by creating a molographic pattern made of a ligand of the membrane protein of interest on the surface of a sensor chip. When living cells are plated onto the sensor chip, the targeted membrane protein aligns to the molographic pattern while still residing in the plasma membrane. Specific molecular interactions between the membrane protein and cytosolic proteins can be measured via the intensity of the molographic focal spot. On the other hand, nonspecific molecular interactions as well as changes to bulk properties of the cells (*e.g.*, mass or volume) and their environment are suppressed. As a result, cell-based molography enables the quantitative analysis of membrane proteins with their endogenous interaction partners in real-time.

In this work, we expand the idea of cell-based molography to interactions between cytosolic proteins and provide a tool for the quantification of virtually any cytosolic PPI in living cells in real-time. This is achieved by synthetically tethering the cytosolic protein of choice (*i.e.*, the bait) to a membrane-spanning linker. The cytosolic portion of this artificial membrane protein serves as a receptor for its endogenous interaction partners (*i.e.*, the prey), while the extracellular portion allows for binding to the template mologram (Fig. 1b). We name this construct a biomimetic membrane receptor or, in short, BioMeR.

In this paper, we present the experimental realization of molography-based cytosolic PPI analysis in living cells using BioMeRs. We rationally designed BioMeRs and optimized their membrane integration by mimicking the structural features of naturally occurring membrane receptors. Then, we exemplified their broad applicability with two model systems: first, by detecting the disruption of an intracellular protein complex through an unlabeled peptide that autonomously diffuses into cells; and second, by quantifying the post-translational modification of endogenous ERK1/2 resulting from upstream MEK inhibition by small molecules.

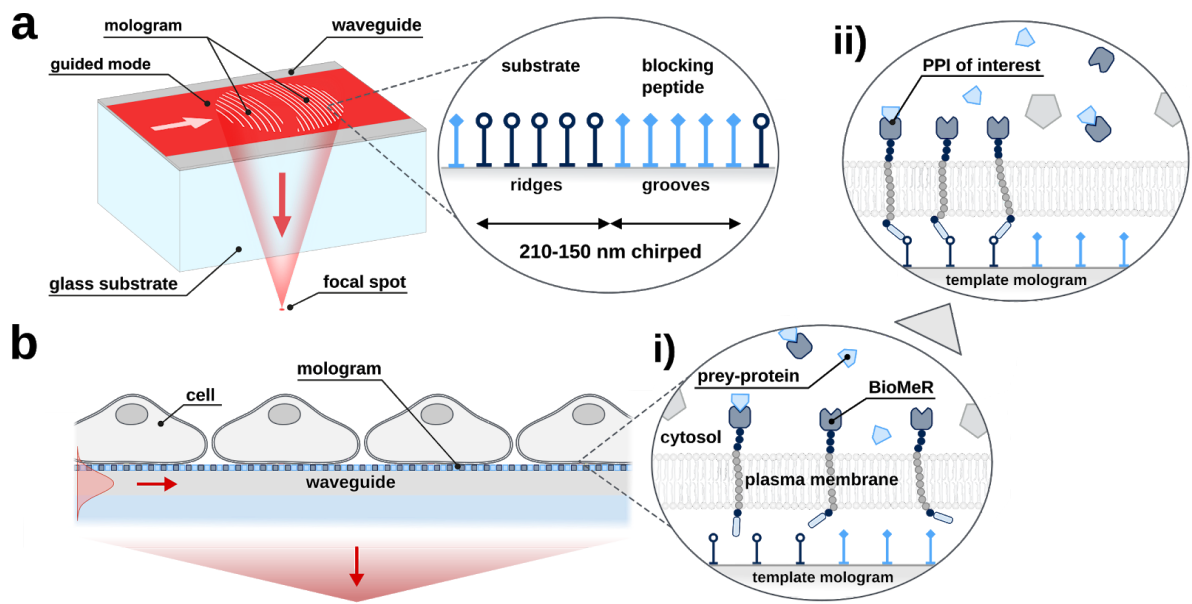

**Figure 1: Principle of focal molography and application of BioMeRs in cell-based molography.**

**a)** A spatially defined 2D nanopattern of molecular binding sites, termed mologram, is generated on a single-mode optical waveguide using reactive immersion lithography (RIL). The ridges of the molographic pattern are functionalized with a substrate that allows for the binding of target molecules whereas the grooves are backfilled with a blocking peptide. Binding of molecules to the mologram results in the physical manifestation of a diffraction grating in the shape of the mologram. When a laser beam is coupled onto the optical waveguide, the molographic grating scatters light at specifically bound molecules so that it constructively interferes in a diffraction-limited focal spot outside of the waveguide. The light intensity in the focal spot is measured to quantify molecular binding in real-time.

**b)** Cells expressing the bait-protein of the PPI to be studied with a BioMeR are plated onto the sensor chip, which is functionalized with a molographic pattern. i) Initially, no molographic signal is detectable as BioMeRs and endogenous proteins freely diffuse within the plasma membrane and the cytosol of cells. ii) Binding of the BioMeRs to the template mologram on the chip via an extracellular tag transfers the molographic pattern to the inside of the cell. Molecular interactions of endogenously expressed cytosolic bait-proteins with BioMeRs can now be quantified in the focal spot of the mologram.

## Results

**Efficient integration of BioMeRs across the cell membrane depends on the signal peptide sequence**

In order to measure cytosolic PPIs by molography, we designed BioMeRs such that one interaction partner of the PPI of interest acts as an intracellular probe that is coherently aligned through the cell membrane by the template mologram. The design mimics structural features of natural transmembrane proteins as follows (Fig. 2a): an N-terminal signal peptide sequence ensures translocation of the newly synthesized BioMeR across the membrane. The signal peptide is followed by a molecular tag, for example an autoreactive SNAP-tag, which is also translocated to the extracellular side and enables straightforward covalent binding to a SNAP-reactive substrate on the template mologram. A single spanning PDGFR-β transmembrane domain is responsible for retaining the protein within the plasma membrane. A flexible linker on the cytosolic side connects the N-terminus of a cytosolic protein to the transmembrane helix and facilitates precessional freedom[25]. Once the protein has been inserted into the membrane, the signal peptide is cleaved off by signal peptidase and is therefore absent in the mature protein[26]. This design ensures the correct orientation of the BioMeR across the plasma membrane, facilitates the alignment to the template mologram and allows to loosely retain the cytosolic protein in the proximity of the membrane. This design can then be genetically encoded in a plasmid (for details refer to Table 1 in the Methods section) that enables the expression of the respective BioMeR by the cell after transfection (Fig. 2b). Since only the correctly oriented BioMeRs, which are residing in the plasma membrane, participate in the molographic assay, their efficient integration into the plasma membrane is crucial for the working principle of the method.

Therefore, from the various signal peptide sequences identified today, we tested different sequences that were previously reported to be highly effective at initiating protein secretion[27–29]. These signal sequences included a modified interleukin-2 (IL-2) signal peptide, a mouse immunoglobulin κ light chain (IgK) peptide and a *Gaussia Luciferase* (Gluc) signal peptide. We employed a construct in which YFP was fused to the C-terminus of the transmembrane helix (termed eYFP-BioMeR), which was transiently expressed in HEK293 cells, and we screened signal peptides for membrane integration efficiency (Fig. 2c). Thereby, the ratio of intra-cellular YFP- to extra-cellular SNAP-tag fluorescence (where the SNAP-tag was coupled to SNAP-reactive surface 649 dye) served as a measure for membrane integration efficiency, which is independent from the overall amount of expressed protein. Out of the three tested signal peptides, we found Gluc to be the most efficient at correctly integrating BioMeRs into the plasma membrane (Fig. 2c), with around 60% of BioMeRs being correctly integrated into the membrane, as well as for the overall protein expression (Fig. 2d). Therefore, *Gluc* was used as the preferred export sequence for constructing all subsequent BioMeRs.

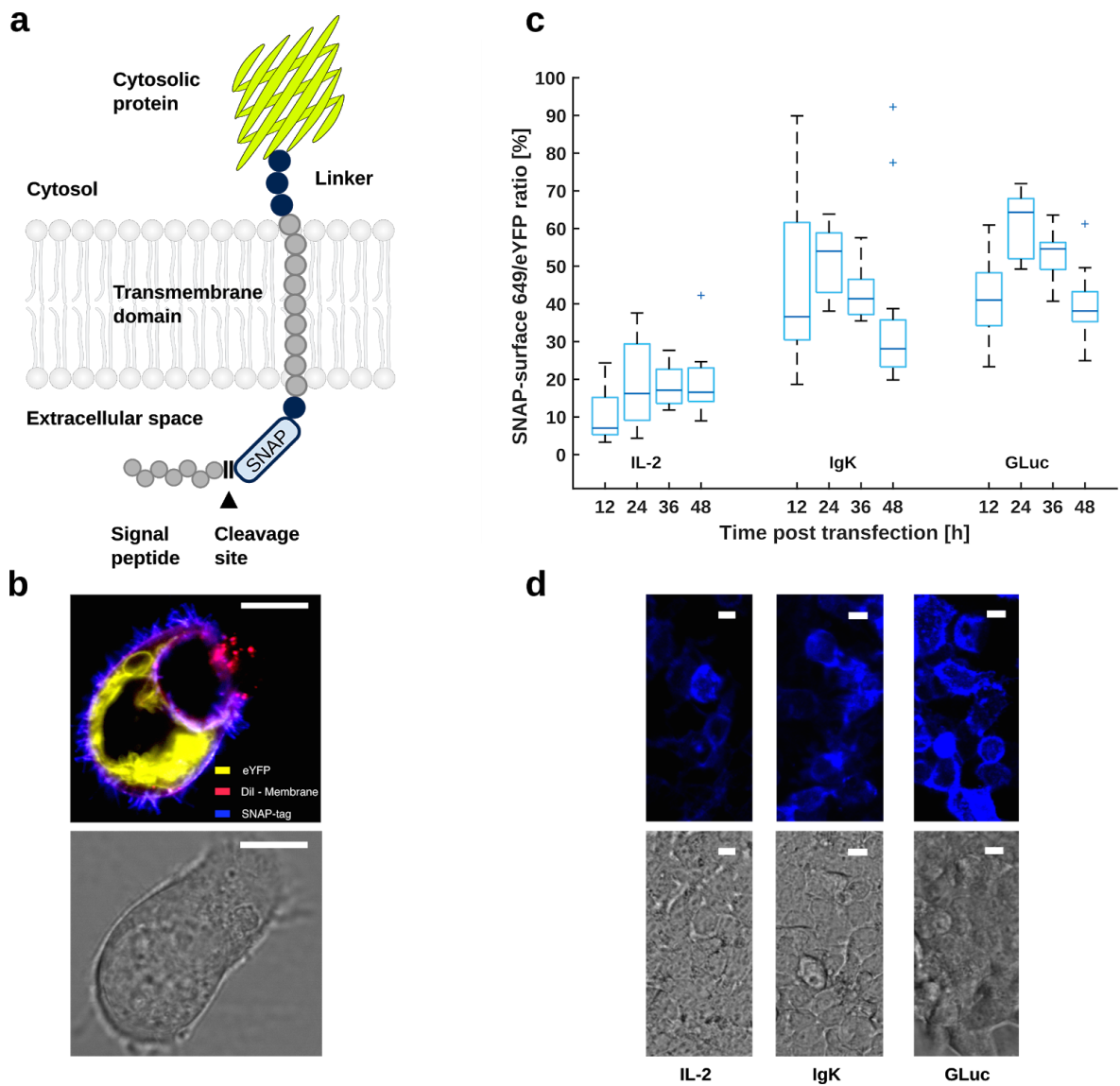

**Figure 2: Integration of BioMeRs across the cell membrane using different signal peptide sequences.**

**a)** BioMeRs consist of an N-terminal cleavable signal peptide that initiates BioMeR translocation across the cell membrane, followed by an extracellular autoreactive SNAP-tag that can bind to a template mologram. A single spanning transmembrane domain connects the SNAP-tag to an intracellular domain that acts as a probe for cytosolic molecules and proteins. The intracellular domain defines the specificity of the BioMeR and can be tailored accordingly: *e.g.*, with a protein or protein domain to monitor specific molecular interactions, with an intracellular antibody (intrabody) or other specificity modules that can be expressed intracellularly; *e.g.,* with DARPins[30] to report the temporal occurrence of specific cytosolic proteins or with an enzyme to report enzymatic activity over time.

**b)** Merged confocal laser scanning microscope (CLSM) (488 nm, 561 nm, 640 nm) and bright field images of membrane-stained HEK293 cells expressing eYFP-BioMeRs. The membrane-impermeable SNAP-reactive surface 649 dye reveals the correct orientation across the cell membrane. The SNAP-tag protein is found on the extracellular side of the plasma membrane and eYFP on the intracellular side. Some BioMeRs are falsely targeted to subcellular locations other than the plasma membrane. Scale bar: 10 µm.

**c)** Staining of the extracellular SNAP-tag with the cell-impermeable dye SNAP-surface 649 reveals membrane-targeting efficiency of IL-2, IgK and *Gaussia Luciferase* signal peptide sequences, reported as a SNAP/eYFP fluorescence ratio. Data represent mean ± s.d. of n = 12 repeats collected over three individual experiments. Outliers are marked with a plus "+" and were due to an increased number of dead cells.

**d)** CLSM and bright field images (640 nm) of SNAP-surface 649 stained HEK293 cells expressing eYFP-BioMeRs with different export sequences 24 h after transfection. Scale bars: 10 µm.

**Cytosolic proteins can serve as artificial receptors for their native interaction partner**

While addressing the lack of small-molecule binding sites in many cytosolic targets, a significant advancement was the understanding that the affinity of PPIs is largely driven by specific residues, called hot regions, as opposed to the entire contact surface of the proteins. Accordingly, most PPIs are mediated by either a linear peptide stretch of one protein interacting with such a hot region on the partner protein (domain-peptide interaction) or a single hot patch on the surface of both interacting proteins (domain-domain interaction)[31,32]. Because of the prominent role of hot segments in PPIs, low molecular weight peptidomimetics have recently been used as drugs for modulating specific interactions[33]. Peptidomimetics are characterized by sufficient conformational flexibility and metabolic stability for translocation across the cell membrane, while providing sufficient affinity to bind to flat PPI interfaces[34].

To demonstrate that BioMeRs are suited to study PPI inhibitors in living cells, we constructed a growth factor receptor-bound protein 2 (Grb2)-BioMeR (Fig. 3a) and expressed it in HEK293 cells. Grb2 is a ubiquitously expressed adaptor protein that takes part in the upstream MAP kinase pathway. For this reason, Grb2 is essential to a variety of basic cellular functions, such as cell cycle progression and cell motility[35]. Amongst others, it links membrane-bound receptor tyrosine kinases (RTKs) to the activation of Ras and its downstream kinases via the guanine nucleotide exchange factor Son of Sevenless 1 (SOS1)[36,37]. The constitutively bound Grb2-SOS1 complex is naturally found close to the cell membrane: therefore, it was chosen as an ideal proof of principle for molographic PPI disruption monitoring.

To this end, we stimulated HEK293 cells stably expressing Grb2-BioMeRs or eYFP-BioMeRs (Fig. 3b) with 50 µM of a previously reported cell-penetrating peptide that blocks the N-terminal SH3- domain of Grb2[38]. Upon stimulation of Grb2-BioMeRs, we observed a rapid decrease in the molographic signal over the course of 25 minutes (Fig. 3b). In comparison, stimulation of eYFP-BioMeRs with the same peptide led to no change in the molographic signal, demonstrating that the peptide evoked a specific response at the Grb2-BioMeR. In cell-based molography, any membrane-bound and cytosolic molecule that is not coherently arranged to the template mologram (*i.e.,* that does not interact with the BioMeR) yields no measurable molographic signal[24]. Therefore, it remains unclear whether the peptide evoked an additional off-target effect on other proteins. Nevertheless, the observed ~30% signal decrease (28.8 pg/mm$^2$) corresponds to a protein that is constitutively bound to Grb2 and displaced by the peptide. Assuming that all Grb2-BioMeRs were occupied by SOS1 which were displaced by the peptide, we can estimate the number and molecular weight of all the coherently arranged proteins. With a molecular weight of 152 kDa, SOS1 would thus account for 74'000 coherently arranged proteins per cell or ~30% of a 500 kDa protein complex where Grb2-BioMeR accounts for approximately 52 kDa. The remaining ~300 kDa may be explained by the presence of RTKs (their molecular weights range from approximately 130-190 kDa and they usually occur as dimers when activated[39]) that are combined with the Grb2-BioMeR.

Even though cell-based molography can quantify the amount of coherently arranged proteins in a cell, it is more difficult to determine the absolute number of a specific protein in a protein complex. This implies that there may be other proteins coherently arranged to the Grb2-BioMeR than the ones suggested. Therefore, for future cell-based molographic systems, we suggest the binding of a molecule with well-known molecular weight or induced unbinding of a portion of a biological complex as a universal method to calibrate the molographic signal in terms of molecular masses of the involved species. Though unbinding of SOS1 was detectable molographically, binding of the low molecular weight peptide (~3 kDa) remained invisible in the molographic signal, which is likely to be due to the insufficient mass resolution of the current setup. The half-maximal effective peptide concentration (EC50) of 13 µM (Fig. 3c) compared reasonably well to the previously reported value of ~15 µM for half maximal proliferation inhibition in cell culture[38].

Compared to RET based assays, BioMeR assisted cell-based molography does not impose constraints on the design of the assay itself (*e.g.,* distance of fluorescent reporters) and requires no modifications on the prey-protein, which is potentially advantageous for assay development. On the other hand, BioMeRs closely confine the bait-protein to the plasma membrane, which could alter the activity of certain proteins.

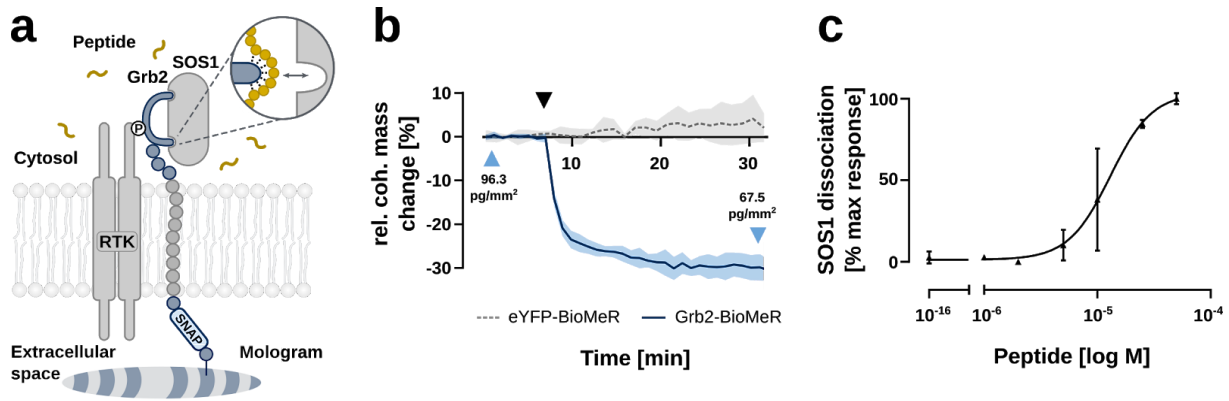

**Figure 3: Grb2-BioMeRs report peptide induced cytosolic PPI disruption in HEK293 cells.**

**a)** Grb2-BioMeRs are covalently bound to a template mologram using an extracellular SNAP-tag. The constitutively bound guanine nucleotide exchange factor SOS1 is then disrupted by a cell-penetrating peptide that binds to the N-terminal SH3 domain of Grb2. Unbinding of SOS1 from the coherently arranged Grb2-BioMeRs can then be observed molographically.

**b)** Stimulation of HEK293 cells expressing Grb2-BioMeRs with 50 µM (black arrow) of the Grb2-SOS1 disruptive peptide leads to a rapid decrease of the molographic signal. In contrast, the molographic signal of cells expressing eYFP-BioMeRs remains unaffected. The figure shows the mean equivalent coherent mass density modulation of 4 individual molograms of one representative experiment.

**c)** Cells expressing Grb2-BioMeRs were treated with increasing concentrations of the disruptive peptide. The half maximal effective peptide concentration after 30 min was found to be approximately 13 ± 2.5 µM. The area under the curve (AUC) was plotted as a function of peptide concentrations. Data represent mean ± s.d. of n ≥ 2 individual experiments.

**Intracellular direct binding assay quantifies post-translational modification of native proteins in real-time**

Post-translational modifications (PTMs) are fundamental regulators of the cell cycle and signal transduction with high pathological relevance[40,41]. Traditionally, PTMs have been studied in fixed cells or lysates with limited temporal resolution. PTMs can neither be directly studied using gene expression analytics nor fluorescent tags, making it notoriously difficult to study PTMs in living cells[42]. For these reasons, binding molecules which are functional in the cytosol and specifically recognize target proteins have gained increasing attention in the past years[30,43]. Besides therapeutic applications, they have also been employed as fluorescent biosensors to image PTMs. However, high background fluorescence or the necessity to also modify the protein under study are major limiting factors of fluorescence-based methods[42]. Cell-based molography is intrinsically insensitive to off-target molecular interactions of the soluble protein, allowing for direct binding assays in living cells. Conversely, off-target interactions of the BioMeR can be tested in control experiments.

Therefore, we designed a specific BioMeR that serves as a cytoplasmically active binding protein to quantify the post-translational phosphorylation of endogenous ERK1/2 in real-time (Fig. 4a). This BioMeR is based on the pE59 DARPin[44], which specifically distinguishes pERK from ERK[44] and was used as a real-time in vivo sensor before[45]. We expressed the pE59-BioMeR in HEK293 cells and diminished ERK1/2 phosphorylation by inhibiting the upstream kinase MEK with the inhibitor PD98509. Initially, phosphorylated ERK1/2 (pERK1/2) is found in an equilibrium state of bound and unbound pERK1/2 with respect to the pE59-BioMeR (Fig. 4b). Upon inhibition of MEK with 20 µM of PD98509, the phosphorylation of ERK is diminished and the intracellular concentration of pERK1/2 is decreasing. The molographic signal from the pE59-BioMeR decreases rapidly upon inhibition of MEK (Fig. 4b, first black arrow) and reaches a plateau after ~30 min, which is in good agreement with the previously reported temporal succession of ERK phosphorylation[46]. In comparison, treatment with an equivalent amount of DMSO as used to dissolve the MEK inhibitor does not lead to a significant drop in the molographic signal. Instead, we observed a slight signal increase after addition of DMSO (Fig. 4b). This increase in signal reported through the pE59-BioMeR may be caused by the mechanical stimulation of the cells upon manual compound addition, which leads to the well-known activation of the MAP kinase pathway or by the decrease in overall refractive index upon dilution of previously DMSO-corrected measurement buffer in the measurement chamber.

To quantify the number of pERK1/2 prior to MEK inhibition, we assume complete dissociation of pERK1/2 from the pE59-BioMeR after 30 min. In this case, the signal drop of ~28 % or 23.7 pg/mm$^2$ corresponds to approximately 215'000 pERK1/2 proteins per cell which were previously bound by the BioMeR. We estimated the half-maximal inhibitory concentration (IC50) of PD98509 to be about 8.2 µM (Fig. 4c), which is in good agreement with the value reported by the manufacturer (7 µM). It is important to note that pE59-BioMeR cells only responded in the first few passages after thawing the cells. Cells that were passaged multiple times gave no reproducible response. We hypothesize that this effect might be due to a change in expression patterns along the MAP kinase pathway induced by the pE59 DARPin.

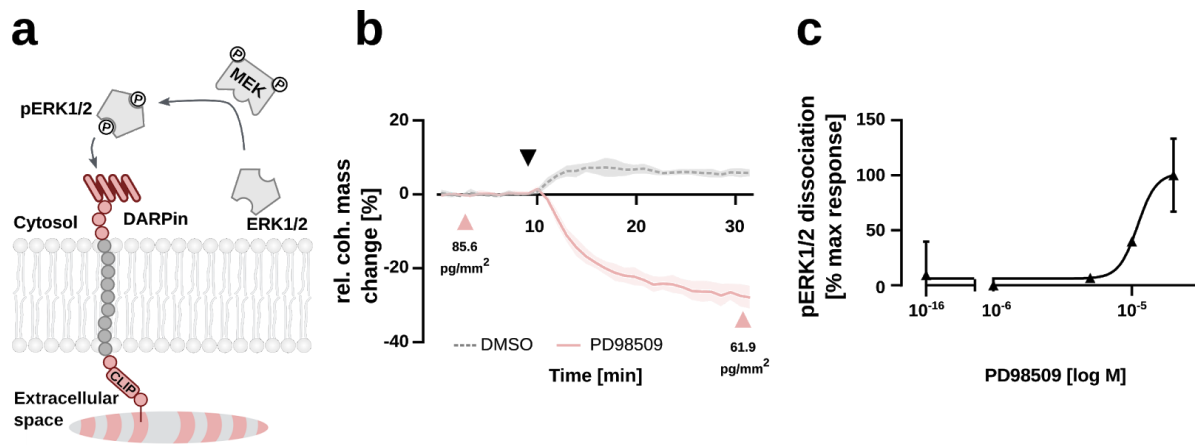

**Figure 4: Label-free monitoring of ERK1/2 phosphorylation in HEK293 cells.**

**a)** pE59-BioMeRs are covalently arranged to a template mologram using an extracellular CLIP-tag. Due to the high affinity and high unbinding rate for pERK1/2, the intracellular pE59 DARPin can accurately track ERK1/2 phosphorylation.

**b)** Inhibition of the upstream kinase MEK with 20 µM of PD98509 (first arrow) decreases the intracellular concentration of pERK1/2, which is detected as a decrease in the molographic signal. The figure shows the mean equivalent coherent mass density modulation of 4 individual mologram of a representative experiment.

**c)** Cells expressing pE59-BioMeRs were treated with increasing concentrations of PD98509. The half maximal effective peptide concentration was estimated to be about 8.2 ± 5.4 µM. The area under the curve (AUC) was plotted as a function of MEK inhibitor concentration. Data represent mean ± s.d. of n ≥ one individual experiments.

## Summary and Outlook

In summary, we have established cytosolic PPI analysis using cell-based molography in living cells. For this purpose, we developed BioMeRs consisting of an extracellular SNAP- or CLIP-tag that can bind covalently to a template mologram, followed by a single span transmembrane domain that connects to a cytosolic bait-protein. When living cells are plated on a molographic sensor, the BioMeRs align to the template mologram and transfer its pattern to the cytosolic space. There, interactions with the bait-protein can be measured. We developed two model systems on the basis of the well-known MAP kinase pathway to demonstrate the working principle of BioMeR-assisted cell-based molography. In the first model system, we disrupted the interaction of Grb2-BioMeRs and SOS1 with a blocking peptide and we quantified the induced dissociation of SOS1 with the molographic setup. In the second model system, we employed a specific DARPin-based BioMeR to monitor the unbinding of phosphorylated ERK1/2 from the DARPin upon upstream kinase inhibition. These models exemplify that BioMeRs can be constructed by endogenous or exogenous proteins to investigate molecular interactions of target proteins.

Due to the label-free signal acquisition of cell-based molography, the protein of interest requires no molecular modifications and is expressed endogenously. Therefore, the design of specific BioMeRs targeting a protein of interest is not restricted by spectral overlaps or by the location of the molecular modification. On the other hand, BioMeR assisted cell-based molography does not allow for interaction analysis with proteins from intracellular locations other than the cytosol and plasma membrane. Moreover, since BioMeRs are covalently linked to a template mologram, subcellular relocalization of proteins upon interaction is spatially restricted. In addition, the cytosolic interaction of interest is artificially brought in proximity of the plasma membrane, which could be unsuitable for certain investigations. Finally, calibration of the molographic signal in terms of molecular masses of the involved species is still unsolved, which makes the evaluation of BioMeR occupation challenging. For these reasons, we foresee BioMeRs to be especially advantageous when derived from cytoplasmically active binding proteins such as DARPins to study PTMs and contribute to PTM-specific drug discovery processes.

# Materials and Methods

Cell culture medium DMEM High Glucose (4.5 g/l) with L-Glutamine (BioConcept, Switzerland), Lipofectamine® 3000, Opti-MEM® I (1X), Versene 1:5000 (1X), Fetal Bovine Serum (FBS), G418, Phosphate-Buffered Saline, pH 7.4 (PBS, Catalog number: 10010-015) and Hank's Balanced Salt Solution with Calcium and Magnesium (HBSS, Catalog number: 14025-050) were purchased from Life Technologies Europe (Zug, Switzerland). TPP 6-well tissue culture plates and tissue culture flasks were from Faust Laborbedarf AG (Schaffhausen, Switzerland). The amine reactive SNAP-tag substrate $O^6$-benzylguanidine (BG-GLA-NHS) and SNAP surface 649 dye were purchased from Bioconcept AG (Allschwil, Switzerland). The amine reactive CLIP-tag substrate $O^6$-benzylcytosine (BC-GLA-NHS) was custom synthesized by KareBay Biochem (Monmouth Junction, NJ, USA). The GRGDSPGSC-(DBCO) peptide was custom synthesized by LifeTein, LLC (Somerset, NJ, USA) whereas the cell-penetrating peptide (KKWKMRRNPFWIKIQRC-CGIRVVDNSPPPPLPPRRRRSAPSPTRV-amide) was custom synthesized by GenicBio Ltd. (Shanghai, China). Azido-PEG4-NHS was obtained from Jena Bioscience (Jena, Germany). PD98509 was purchased from Tocris Bioscience (Bristol, UK). All other chemicals were purchased from Sigma-Aldrich Chemie GmbH (Buchs SG, Switzerland). The PAA-g-PEG-NH-PhSNPPOC copolymer, used as a biocompatible coating, was a kind gift of Roche (Basel, Switzerland). Zeptosens thin-film optical waveguides with a 145 nm $Ta_2O_5$ layer with the in- and out- coupling gratings covered with a 1 μm thick layer of $SiO_2$ by IMT Masken und Teilungen AG (Greifensee, Switzerland) were a kind gift of Roche (Basel, Switzerland).

## Cell culture

HEK293 wild type cells were cultured in DMEM medium containing 10%-v/v fetal bovine serum at 37°C in a cell incubator with 5% $CO_2$. Cells stably expressing Grb2- and pE59-BioMeRs were cultured in medium additionally supplemented with 600 μg/ml G418.

## Stable cell line generation

For the generation of transiently expressing eYFP-BioMeRs cells, HEK293 cells were seeded in 6 well plates at 600-800k cells/well in 2 ml culture media and transfected using Lipofectamine 3000 transfection reagent according to the manufacturer's protocol. In order to establish stable Grb2- and pE59-BioMeR cell lines, transiently transfected cells were grown in medium supplemented with 1 mg/ml G418 for approximately 20 days. Afterwards, neomycin-resistant cells were stained using a SNAP- or CLIP-surface 649 dye and selected by flow cytometry.

**Plasmids**

Plasmids encoding for different BioMeRs were purchased from Invitrogen GeneArt Gene Synthesis service by Thermo Fisher Scientific. All synthetic genes were assembled from synthetic oligonucleotides and/or PCR products and inserted into a pcDNA3.1(+) vector backbone by the manufacturer. The plasmid DNA was purified from bacteria, the concentration was determined by UV spectroscopy and the final constructs were verified by sequencing by the manufacturer. Plasmids were delivered in TE buffer (10 mM Tris-HCl, pH 8.0, 1 mM EDTA) at a concentration of 1mg/ml and stored at 4°C. The amino acid sequences of the individual BioMeR subdomains are displayed in Table 1.

| Feature | Sequence (N to C-terminus) |
| --- | --- |
| IL-2 signal peptide | MRMQLLLLIALSLALVTNS |
| IgK signal peptide | METDTLLLWVLLLWVPGSTGD |
| *Gaussia luciferase* signal peptide | MGVKVLFALICIAVAEA |
| SNAP-tag | MDKDCEMKRTTLDSPLGKLELSGCEQGLHRIIFLGKGTSAADAVEVPAPAAVLGGPEPLMQATAWLNAYFHQPEAIEEFPVPALHHPVFQQESFTRQVLWKLLKVVKFGEVISYSHLAALAGNPAATAAVKTALSGNPVPILIPCHRVVQGDLDVGGYEGGLAVKEWLLAHEGHRLGKPGLG |
| CLIP-tag | MDKDCEMKRTTLDSPLGKLELSGCEQGLHRIIFLGKGTSAADAVEVPAPAAVLGGPEPLIQATAWLNAYFHQPEAIEEFPVPALHHPVFQQESFTRQVLWKLLKVVKFGEVISESHLAALVGNPAATAAVNTALDGNPVPILIPCHRVVQGDSDVGPYLGGLAVKEWLLAHEGHRLGKPGLG |
| PDGFR-β Transmembrane domain | AVGQDTQEVIVVPHSLPFKVVVISAILALVVLTIISLIILIMLWQKKPR |
| Linker | GGGGSGGGGSGSAGSAAGSGEFGGGGSGGGGSG |
| eYFP | MVSKGEELFTGVVPILVELDGDVNGHKFSVSGEGEGDATYGKLTLKFICTTGKLPVPWPTLVTTFGYGLQCFARYPDHMKQHDFFKSAMPEGYVQERTIFFKDDGNYKTRAEVKFEGDTLVNRIELKGIDFKEDGNILGHKLEYNYNSHNVYIMADKQKNGIKVNFKIRHNIEDGSVQLADHYQQNTPIGDGPVLLPDNHYLSYQSALSKDPNEKRDHMVLLEFVTAAGITLGMDELYK |
| Grb2 | MEAIAKYDFKATADDELSFKRGDILKVLNEECDQNWYKAELNGKDGFIPKNYIEMKPHPWFFGKIPRAKAEEMLSKQRHDGAFLIRESESAPGDFSLSVKFGNDVQHFKVLRDGAGKYFLWVVKFNSLNELVDYHRSTSVSRNQQIFLRDIEQVPQQPTYVQALFDFDPQEDGELGFRRGDFIHVMDNSDPNWWKGACHG |

|  | QTGMFPRNYVTPVNRNV |
|---|---|
| pE59 DARPin | DLGKKLLEAARAGQDDEVRILMANGADVNALDEDGLTPLHL AAQLGHLEIVEVLLKYGADVNAEDNFGITPLHLAAIRGHLEIV EVLLKHGADVNAQDKFGKTAFDISIDNGNEDLAEILQ |

**Table 1: Amino acid sequences of individual BioMeR subdomains.**

**Preparation of sensor chips**

Thin-film optical waveguides were treated using a protocol reported previously[22]. Briefly, waveguides were washed with 0.1% aqueous Tween-20, followed by ultrasound-assisted washing in MilliQ water, isopropanol and toluene. The chips were then immersed in warm Hellmanex III for 1 min and thoroughly rinsed with MilliQ water. Afterwards, the chips were cleaned in ultrasound-assisted highly oxidizing Piranha solution ($H_2SO_4/H_2O_2$ 7:3) for 30 min and thoroughly washed with MilliQ water. Chips were centrifuge-dried at 800 rcf for 2 min and activated by oxygen plasma. After plasma treatment, the chips were immediately immersed in the PAA-g-PEG-NH-PhSNPPOC graft copolymer coating solution (0.1 mg/ml in 1 mM HEPES pH 7.4) for 60 min for copolymer adlayer formation. To fully passivate the layer, the chips were washed with MilliQ water and ethanol and immersed in a 25 mM solution of methyl chloroformate in anhydrous acetonitrile containing 2 equiv. of N,N-diisopropylethylamine for 5 min. The coated chips were washed with ethanol and MilliQ water and blow-dried by a nitrogen jet. Prepared sensor chips were stored in the dark at 4 °C until further use.

**Preparation of template molograms**

Molograms were prepared following the standard reaction immersion lithography process described previously[22,47]. In short, a previously copolymer-coated sensor chip was placed in a custom holder. The phase mask used to generate the molograms was aligned using an alignment help and the gap between the chip and phase mask was filled with a solution of 0.1 %-v/v hydroxylamine in DMSO. Then, a photolithographic exposure was conducted at 405 nm with a dose of 2000 mJ/cm² in a custom-built setup to create the molographic pattern. After illumination, the chip was washed with isopropanol and MilliQ water and the deprotected ridges were functionalized with 1 mM amine reactive SNAP- or CLIP-tag substrate (respectively BG-GLA-NHS or BC-GLA-NHS), to which the SNAP- or CLIP-tag protein can covalently bind. To facilitate cell adhesion to the chip, remaining PhSNPPOC groups were removed by flood exposure and the free binding sites were functionalized with the hetero-bifunctional crosslinker azido-PEG4-NHS, to which the extracellular matrix-mimicking peptide GRGDSPGSC-(DBCO) was coupled. The setup and phase mask used to generate template molograms was a kind gift of Roche (Basel, Switzerland).

**Molographic cell measurements**

Cells stably expressing Grb2- or pE59-BioMeRs were grown to 60-80% confluency in T25 culture flasks. Cells that were stimulated with the cell-penetrating peptide were starved in FBS-free media overnight. Cells were washed twice with warm PBS, incubated with 1x Versene for 5 min and gently resuspended in warm cell culture medium. The cells were then centrifuged at 50 rcf for 1 min and resuspended in media two times sequentially to minimize cellular debris. Cells were seeded on the molographic chip in an incubation chamber containing 500 µl cell culture media. Note, the cells were only seeded when their viability exceeded 90%, as determined by a Countess automated cell counter (Invitrogen). The cells were kept in a $CO_2$ incubator at 37 °C for 2 h to allow cell adherence to the sensor chip (and covalent interaction of the SNAP- or CLIP-tag of the BioMeRs with the substrate on the chip). The incubation chamber containing the cells was then washed twice with warm HBSS buffer (supplemented with 20 mM HEPES, pH 7.4), adjusted for DMSO if needed and transferred to a modified F3000 ZeptoReader (Zeptosens), which was kept at 35 °C. The molographic chip was allowed to temperature-equilibrate inside the ZeptoReader for 5-10 min before performing the assay. For all molographic assays, the signal was monitored for 7-10 min (baseline measurement) before addition of the respective compound. Typical instrument parameters for molographic signal acquisition were as follows: one image was acquired every 10 s using the 635 nm laser with an integration time of 0.25-1 s depending on the intensity of the initial signal and a grey filter value of 0.001 in the illumination path of the ZeptoReader. For the evaluation of the molographic signal, automation (AutoHotkey) and evaluation (MATLAB) scripts were used[22].

**Data analysis and calculations**

As described previously[24], the equivalent coherent mass modulation was obtained via an algorithm that computes the power in the waveguide from the scattered background intensity of the waveguide mode as described in ref. [23]. The anisotropy of the scattering was assumed to be 0.05418. The damping constant was computed separately for every chip and mologram field (3-6 dB/cm) and the numerical aperture of the Zeptoreader is 0.33. Because the Zeptoreader is incapable of resolving the Airy disk, the measured signal (arb. units) was divided by the expected Airy disk area to obtain a quantity that was proportional to the average intensity in the Airy disk. The quantity proportional to the background intensity was computed from the background signal (arb. units) by dividing it by the pixel size of the ZeptoReader (12.5 x 12.5 µm). The ratio of these two quantities was then used in Eq. 11 of ref. [23] with the necessary scaling factors to obtain the equivalent coherent mass density from the average intensity in the Airy disk. The algorithm was implemented in MATLAB as well as in Python.

The number of proteins per cell was derived as follows. The equivalent coherent mass density was multiplied with the area of the mologram footprint to receive the coherent mass. We assumed that the mologram with a diameter of 400 µm was covered by about 1000 cells. However, due to the central curved recess area BioMeRs are only aligned in about 80% of the cells to the template mologram. The coherent mass was thus divided by 800 cells to receive the coherent mass per cell. To account for imperfect alignment of the BioMeRs, the coherent mass per cell has to be divided by the analyte efficiency (0.24)[47] to receive the total mass of the proteins per cell. Dividing with the molecular weight of the protein and multiplying with Avogadro constant, then yields the number of proteins per cell.

**CLSM imaging and segmentation**

For fluorescence imaging, cells were seeded on a 24-glass bottom well plate at 50% confluence and transfected after 24h as described previously. The transfection medium was replaced after 12 h with complete medium. Cells were imaged 12, 24, 36 and 48 hours after transfection using an Olympus FluoView FV3000 confocal laser scanning microscope. Prior to imaging, cells were incubated with SNAP-Surface 649 dye for 30 mins and then washed three times with warm PBS. During imaging, cells were kept at 37 °C with 5% $CO_2$. The eYFP and SNAP-Surface 649 channels were acquired simultaneously with a 20x objective using 514 nm excitation / 527 nm emission wavelengths and 651 nm excitation / 667 nm emission wavelengths for the green and red channels respectively. CLSM settings were kept the same for all samples.

Acquired images were batch converted to .tiff format using Fiji[48] and imported into Matlab for further processing. Images were firstly split into three different channels (bright field, green and red). For the green and red channels, a normalized histogram of intensities was plotted using the grayscale values of each channel. The optimal threshold to distinguish between object and background pixels was calculated for each image using the following equation:

$$T = \frac{\mu_o + \mu_b}{2} + \frac{\sigma^2}{\mu_o - \mu_b} ln\left(\frac{P_b}{P_o}\right)$$

where $\mu_o$ and $\mu_b$ are the mean values of the normal distributions of object pixels and background pixels respectively, $P_o$ is the probability that a pixel belongs to the object and $P_b$ the probability that it belongs to the background. The above equation assumes that the variances of the Gaussian curves of object and background pixels are equal ($\sigma_o = \sigma_b = \sigma$). The total amount of fluorescence was then calculated by summing the intensities over all object pixels in each channel. The ratio of SNAP-Surface 649 (red channel) to eYFP fluorescence (green channel) was calculated by dividing the total red and green fluorescence for each image. The value of total eYFP fluorescence and the ratio between SNAP-Surface 649 and eYFP fluorescence were then plotted as a box plot for GLuc, IL-2, IgK and RV-GLuc samples at 12 h, 24 h, 36 h and 48 h.

## Acknowledgments

We acknowledge the Roche Innovation Center in Basel, and Christof Fattinger in particular, for providing the molographic chips and the experimental setup that we used for synthesis of molograms by reactive immersion lithography. Furthermore, we thank Prof. Paola Picotti (ETH), Prof. Matthias Peter (ETH) and Prof. Olivier Pertz (University of Bern) as well as Dr. Maria Waldhoer (Interax), Dr. Maciej Dobrzynski (University of Bern), Dr. Nako Nakatsuka (ETH), Dr. Stephanie Hwu (ETH) and Aline F Renz (ETH) for helpful discussions on the topic. Finally, we thank Linda Molli (University of Pisa) for proofreading the manuscript.

## Competing Interest Statement

A.M.R., I.I., Y.B., I.L., F.T. and J.V. declare that they have no competing financial interests. A.F. is involved with the commercialization of focal molography. A.P. is involved with the commercialization of the DARPin technology.